\documentclass[aps,prl]{revtex4}
\usepackage[intlimits]{amsmath}
\usepackage{amsfonts,graphicx}
\usepackage{mciteplus}

\newcommand{\bra}[1]{\left\langle #1 \right|}
\newcommand{\ket}[1]{\left| #1 \right\rangle}
\newcommand{\xx}{\mathbf{x}}
\newcommand{\aver}[1]{\left\langle #1 \right\rangle}

\newcommand{\be}{\begin{equation}}
\newcommand{\ee}{\end{equation}}
\newcommand{\bea}{\begin{eqnarray}}
\newcommand{\eea}{\end{eqnarray}}
\newcommand{\beqa}{\begin{eqnarray*}}
\newcommand{\eeqa}{\end{eqnarray*}}
\newcommand{\nn}{\nonumber}

\begin{document}

\title{Entangling photons through nonlinear response of quantum wells to ultrashort pulses}

\author{Mikhail Erementchouk}
\author{Michael N. Leuenberger}\email{mleuenbe@mail.ucf.edu}
\affiliation{NanoScience Technology Center and Department of Physics, University of Central
Florida, Orlando, FL 32826}

\begin{abstract}
We show that many-body correlations among excitons originating from the Pauli exclusion principle
in a quantum well embedded inside a microcavity
provide a possibility to produce pairs of entangled photons by ultrashort laser pulses with a yield of $\sim 10^{-2}$.
The quantum-field theoretical two-particle density matrix in second quantization
is used to calculate entanglement for arbitrary emission angles.
Largest response can be expected at symmetric emission angles for
resonances with the heavy-heavy and light-light two-exciton states with
remarkably nontrivial dependence of entanglement on the emission angles and
on the ellipticity parameters of the excitation. We show that the angle dependence
can be tailored by means of the microcavity. Interestingly, the emitted
entangled 2-photon states are always in a triplet state.
\end{abstract}

\maketitle

Current entangled-photon sources are mainly based on the parametric
down-conversion (PDC) inside nonlinear crystals\cite{Shih,Ou}, such as
BBO crystals\cite{Altepeter,Kwiat}. These sources suffer from two serious
limitations. First, since the far off-resonant three-photon scattering
contains two far off-resonant virtual states, the entangled-photon
production yield is very low\cite{Bouwmeester,Mandel}, which limits the
brightness of entangled-photon sources based on nonlinear crystals, leading
to low signal-to-noise ratios and long measurement times\cite{Tanzilli}.
Second, PDC produces entangled photons with the twice as long wavelength as
the pump photons, which limits the operating wavelength\cite{Bouwmeester}.
Therefore, the problem of alternative sources of entangled photons is of the
great importance.
Quantum
dot (QD) structures have already been used for
this\cite{Hafenbrak,STEVENSON:2006:ID3501,Akopian2} making use of the
relaxation of two excitons into one bound biexciton on a QD, although QD
structures cannot achieve the brightness of quantum well (QW) structures.


The current limitation of QD structures is the low operating
temperature of about 4 to 30 K, which is mainly due to the
decoherence arising from exciton-phonon and hyperfine interactions.
QW structures face the additional decoherence source due to Coulomb
interactions, even at low temperatures. We therefore investigate the
possibility to use the short time response
for the production of entangled photons at time scales for which the
biexciton binding energy $E_{xx}$ cannot be resolved anymore, i.e.
$T\ll\hbar/E_{xx}$, and therefore the Coulomb interactions is negligible.
In order for this method to be effective, a microcavity is required to
extract quickly the entangled photons. In addition, the microcavity can be
used to tailor the angle dependence of the entanglement. We show that the
Pauli exclusion principle, which is instantaneous, is sufficient to produce
entangled photons from a QW structure with a high yield of $10^{-2}$. This
method of producing entangled photons is radically different from the
well-known method based on the bound biexciton state
\cite{STEVENSON:2006:ID3501,Young,Hafenbrak,Shields,EDAMATSU:2004:ID3421,Oohata,SAVASTA:2003_SST}.



The excitation of a QW in a cavity by the external field and emission of the
photons due to the radiative recombination are driven by the interaction of
the QW with the photonic modes of the cavity. They are found by
quantizing the electromagnetic (EM) field in the whole space while taking into
account the one-dimensional (1D) spatial modulation of the refractive index
$n(z)$ with $z$ axis coinciding with the crystal growth direction. The
states of the EM field are specified by $\widehat{k} =
(\mathbf{k}_\|, \omega, h)$. Here $\mathbf{k}_\|$ is the in-plane
wavevector, $\omega$ is the frequency, $h = s,p$ denotes the polarization
state. In units $\hbar = 1$, $c=1$ we
present the quantized field as (see e.g. Ref.~\onlinecite{KIRA:1999})
\begin{equation}\label{eq:field_quantized}
  \mathbf{A} = \frac{1}{(2\pi)^{3/2}}\sum_{\widehat{k}} {\boldsymbol{\epsilon}}_{\widehat{k}}(z) \frac{1}{\sqrt{2 \omega_{\widehat{k}}}}
  u_{\widehat{k}}(z)e^{i \mathbf{k}_\|\cdot\boldsymbol{\rho}} a^\dagger_{\widehat{k}} + \text{h.c.},
\end{equation}
where $\boldsymbol{\mathbf{\epsilon}}_{\widehat{k}}$ are the unit
polarization vectors ($\boldsymbol{\mathbf{\epsilon}}_s$ lies in the plane
of QW), $u_{\widehat{k}}(z)$ is the spatial distribution of the field along
the $z$-axis found as the solution of the respective 1D scattering problem,
$\boldsymbol{\rho}$ is the coordinate in $(x,y)$ plane and
$a^\dagger_{\widehat{k}}$ is the photon creation operator. The summation
over $\widehat{k}$ implies the integration over the continuous quantum
numbers and summation over the discrete ones.

Entanglement of the photons produced in the course of the radiative
relaxation of the pumped semiconductor is found considering the two-photon
density matrix
\begin{equation}\label{eq:def_two-photon_density}
  \rho_{\widehat{q}_1,\widehat{q}_2}^{\widehat{k}_1,\widehat{k}_2}(t) =
  \bra{\Psi(t)}a^\dagger_{\widehat{q}_1} a^\dagger_{\widehat{q}_2} a_{\widehat{k}_1} a_{\widehat{k}_2}\ket{\Psi(t)},
\end{equation}
where $\ket{\Psi(t)}$ is the state of the semiconductor-photon system. In lowest-order perturbation theory non-vanishing terms
result from the contribution of the two-photon states
into $\ket{\Psi(t)}$, leading to
\begin{equation}\label{eq:two_photon_density_factor}
  \rho_{\widehat{q}_1,\widehat{q}_2}^{\widehat{k}_1,\widehat{k}_2}(t) =
  \Psi^*_{\widehat{q}_1,\widehat{q}_2}(t)
  \Psi_{\widehat{k}_1,\widehat{k}_2}(t)
\end{equation}
with $\Psi_{\widehat{k}_1,\widehat{k}_2}(t) = \bra{0}
  a_{\widehat{k}_1} a_{\widehat{k}_2} \ket{\Psi(t)}$,
where $\ket{0}$ is vacuum of the combined system.
The free dynamics in the interaction picture is given by
\begin{equation}\label{eq:two_photon_psi_interaction}
  \Psi_{\widehat{k}_1,\widehat{k}_2}(t) = \bra{0}
  a_{\widehat{k}_1}(t) a_{\widehat{k}_2}(t) \mathcal{T}_+ \exp\left\lbrace-i
  \int_0^t dt' \widetilde{H}_{int}(t')\right\rbrace \ket{0},
\end{equation}
where $a_{\widehat{k}}(t)$ and $\widetilde{H}_{int}(t)$ are the photon
annihilation operator and the Hamiltonian of the light-matter interaction,
respectively, and $\mathcal{T}_+$ is the
time-ordering operator.

The low energy excitations are conveniently accounted by introducing
the exciton operators according to $\ket{\mu} = B_\mu^\dagger \ket{0}$,
where $\ket{\mu}$ is the hole-electron pair state either bound or unbound
corresponding to energy $E_\mu$, i.e. $H_{SC}\ket{\mu} = E_\mu \ket{\mu}$
with $H_{SC}$ being the Hamiltonian of the nonperturbed semiconductor. For
simplicity we assume that the QW can be approximated by a 2D plane situated
at $z=z_0$. In this case the exciton states are characterized by the spin
states of the hole and the electron constituting the pair, the center of
mass momentum in the plane of the well, $\mathbf{K}$, and other quantum
numbers, $n_\mu$, so that $\ket{\mu} = \ket{\sigma_\mu, s_\mu,
\mathbf{K}_\mu, n_\mu}$. Denoting by $\phi_\mu(\xx, \xx')$ the exciton
(hole-electron) wave function corresponding to the state $\mu$ we represent
the exciton operator as $B_\mu = \int d\xx \, d\xx'\, \phi^*_\mu(\xx,
\xx')c_{s_\mu}(\xx')v_{\sigma_\mu}(\xx)$,
where $c_s$ and $v_\sigma$ are the annihilation operators for the electrons
with the spin $s$ and the holes with the spin $\sigma$, respectively. In
terms of the exciton operators the light-matter interaction Hamiltonian is
$H_{int} = \sum_\mu \left(\bar{\mathcal{A}}_\mu B_\mu + \mathcal{A}_\mu
B^\dagger_\mu \right)$ with ${\mathcal{A}}_\mu$ and $\bar{\mathcal{A}}_\mu$
defined as convolutions of $\mathbf{A}(\xx)$ with
${\mathbf{d}}_{\sigma_\mu, s_\mu} \phi_\mu(\xx,\xx)$ and its conjugate,
respectively. Here $\mathbf{d}_{nn'} = e/m \bra{n} \mathbf{p} \ket{n'}$ with
$\bra{n} \mathbf{p} \ket{n'}$ is the matrix element of the momentum
operator between the bands $n$ and $n'$.

In order to distinguish between the processes of the excitation of the
semiconductor and the radiative exciton recombination we separate the
contributions of the quantized field of emitted photons,
$\mathbf{A}^{(q)}(\xx)$, and of the classical pumping field,
$\mathbf{A}^{(cl)}(\xx)$, into the external field $\mathbf{A}(\xx) =
\mathbf{A}^{(q)}(\xx) + \mathbf{A}^{(cl)}(\xx)$.
In lowest-order perturbation theory we can neglect the processes
of re-emission and reabsorption,
thus obtaining
\begin{equation}\label{eq:interaction_approximated}
  H_{int}
\approx \sum_\mu \left(\bar{\mathcal{A}}^{(q)}_\mu B_\mu + \mathcal{A}^{(cl)}_\mu B^\dagger_\mu \right),
\end{equation}
which we use in Eq.~\eqref{eq:two_photon_psi_interaction}. Since in the
low-intensity regime the higher-order terms of the light-matter interaction
that affect the form of the photonic modes in the expression for the density
matrix are negligible, the pump field is found by solving the respective
initial value problem for the cavity alone without the QW. From now on we will omit the upper index for the quantized
field.

Expanding the exponential term in Eq.~\eqref{eq:two_photon_psi_interaction}
we obtain various terms depending on ordering of the operators $B$ and
$B^\dagger$. We leave only the terms describing the response along the
directions different from the direction of the incident excitation field,
where it is not blurred by the non-scattered field and by the linear
(single-photon) response. Using the $\delta$-functional approximation for
the $z$ dependence of the exciton wave function the two-photon amplitude of the outgoing photons can
be presented as
\begin{equation}\label{eq:two_particle_polarization}
 \Psi_{\widehat{k}_1,\widehat{k}_2}(t) = u_{\widehat{k}_1}(z_0)u_{\widehat{k}_2}(z_0)
 \boldsymbol{\epsilon}_{\widehat{k}_1}(z_0)\cdot \overleftrightarrow{M}_{\mathbf{k}_1,\mathbf{k}_2}(t)\cdot
 \boldsymbol{\epsilon}_{\widehat{k}_2}(z_0),
\end{equation}
where the tensor $\overleftrightarrow{M}_{\mathbf{k}_1,\mathbf{k}_2}(t)$ depends only on
the direction of propagation of the outgoing photons but neither on their
polarizations nor on the choice of the solutions of the scattering problems
[see the supplement for the explicit form of
$\overleftrightarrow{M}_{\mathbf{k}_1,\mathbf{k}_2}(t)$]. This information and the
effect of the photonic density of states modified by the cavity are
contained in the amplitudes $u_{\widehat{k}_{1,2}}(z_0)$. If the field
distribution inside the cavity has the maximum near $z_0$ this results in
accordingly amplified two-particle amplitudes
$\Psi_{\widehat{k}_1,\widehat{k}_2}$. On the contrary, if for one
polarization, $s$ or $p$, the amplitude is significantly smaller comparing
to the other this can be easily shown to imply significant decrease of the
photon entanglement and so on.

The important property of the two-photon amplitudes follows from the
conservation of the total in-plane momentum. Assuming the normal incidence
of the excitation field this leads to the important restriction on the
two-photon amplitudes $\Psi_{\widehat{k}_1,\widehat{k}_2} \propto
\delta(\mathbf{k}_{1, \|} + \mathbf{k}_{2, \|})$. Since the QW is invariant
only with respect to in-plane translations the restrictions is imposed only
on the in-plane component of the wave vectors of the outgoing photons.


As Eq.~\eqref{eq:two_photon_density_factor} shows, the perturbation theory
produces the density matrix corresponding to a pure state of the two photon
system. Thus we can directly apply the standard machinery for evaluating
entanglement of two photons as the von Neumann entropy of the reduced
density matrix.
The dependence of entanglement on the direction of propagation of the
outgoing photons is very complex.
In order to consider the main features of entanglement of the emitted
photons, we discuss the details of the case of a single QW without the cavity,
leaving the details of the case including cavity to another paper.
In the case without cavity one obtains a compact expression for the reduced
single-photon density matrix
\begin{equation}\label{eq:reduced_density_matrix}
  \rho_{\boldsymbol{\epsilon}}^{\boldsymbol{\epsilon}'}(t;\mathbf{k}_1,\mathbf{k}_2) =
\boldsymbol{\epsilon} \cdot\overleftrightarrow{M}_{\mathbf{k}_1,\mathbf{k}_2}(t)
(\overleftrightarrow{1}- \widehat{\mathbf{e}}_2 \otimes \widehat{\mathbf{e}}_2)
\overleftrightarrow{M}_{\mathbf{k}_1,\mathbf{k}_2}^\dagger(t) \cdot
\boldsymbol{\epsilon}',
\end{equation}
where $\overleftrightarrow{1}$ is the unit tensor and the argument of
$\rho_{\boldsymbol{\epsilon}}^{\boldsymbol{\epsilon}'}$ shows the dependence
of the reduced density matrix on the wave vectors of the \emph{pair} of the
outgoing photons. The eigenvalues of the reduced density matrix determine
entanglement as $E_N = -\widetilde{\rho}_1 \log_2(\widetilde{\rho}_1)
-\widetilde{\rho}_2 \log_2(\widetilde{\rho}_2)$ with $\widetilde{\rho}_{1,2}
= \rho_{1,2}/(\rho_1 + \rho_2)$.
There are two terms in
$\overleftrightarrow{M}_{\mathbf{k}_1,\mathbf{k}_2}(t)$, which describe different
physical origins of entanglement of the emitted photons. One term describes
the creation of entanglement due to the Pauli exclusion principle while
another term accounts for the effect of the Coulomb interaction.
Staying in the ultrashort time limit, we can neglect the effect of the Coulomb interaction.

\begin{widetext}

\begin{figure}[h]
  \includegraphics[width=6.5in]{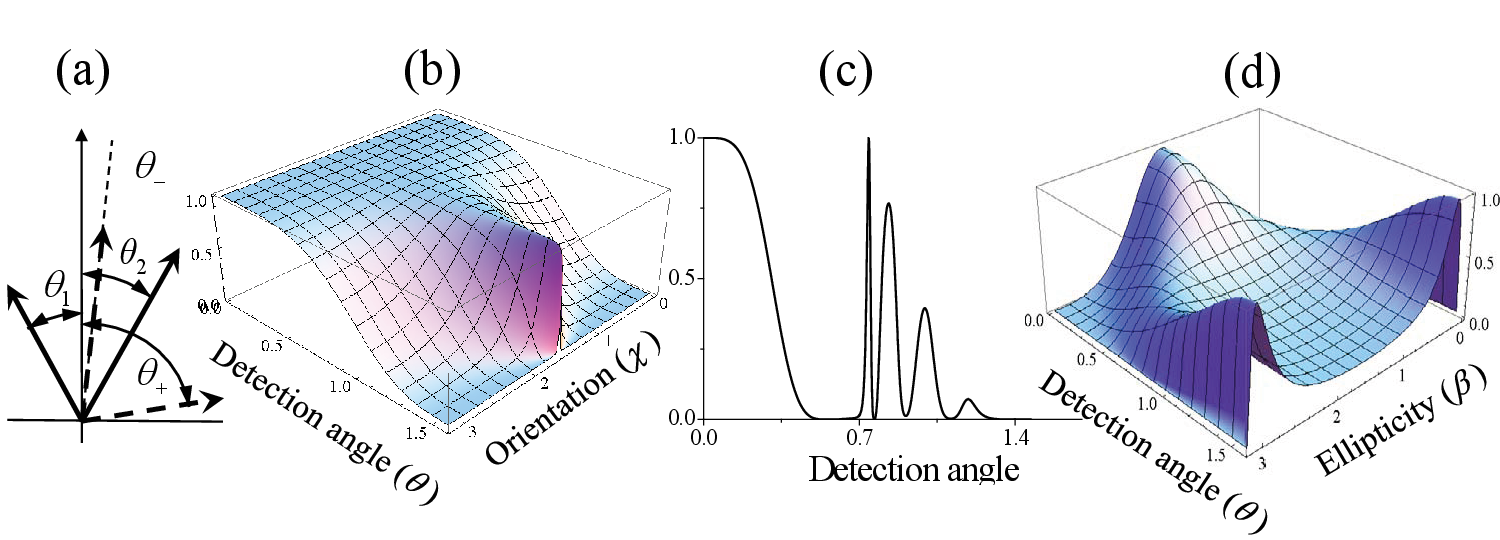}\\
  \caption{(a) The emission directions. The symmetric case corresponds to the
  heavy-heavy and light-light two-exciton resonances. The asymmetric directions
  $\theta_2 = \theta_\pm$ correspond to the heavy-light resonance. (b) -- (d) The Dependence
  of entanglement (vertical axes, scale from $0$ to $1$). (b) $E_N(\theta, \beta = \pi/2, \chi)$
  in the vicinity of the heavy-hole exciton resonance. (c) $E_N(\theta, \beta = \pi/2, \chi = 0)$
  in the presence of the cavity with 20 layers of the same optical width at $\theta = 0$, the
  index contrast is $1.5$, the heavy-hole exciton resonance is tuned at the middle of the first
  stop-band at $\theta = 0$. (d) $E_N(\theta, \beta, \chi = 0)$ near the light-hole
  exciton resonance.
%
  }\label{fig:symm_ent}
\end{figure}

\end{widetext}

{\it For the resonance with heavy-hole excitons} the non-normalized
eigenvalues of the single-particle density matrix are proportional to the
solutions of
\bea\label{eq:eigen_values_symmetric}
 & &  \rho^2
  - \frac \rho 2 \left\{\cos(2\chi)\sin^2(\beta)\sin^4(\theta)+
        \left[1+\cos^2(\beta)\right] \right. \nn\\
& & \times\left.\left[1+\cos^2(\theta)\right]^2\right\}
 + \sin^4(\beta)\cos^4(\theta) = 0,
\eea
where $\beta$ and $\chi$ are the polar and azimuthal angles on the Poincare
sphere describing the polarization state of the excitation field,
\cite{Shurkliff} $A_+=e^{i\chi/2}\cos(\beta/2)$ and
$A_-=e^{-i\chi/2}\sin(\beta/2)$. So that $\beta=0,\pi, \pi/2$ correspond to
the left and right circular  and linear polarizations, respectively, and
$\chi/2$ is the angle between the axis of the ellipse of polarization and
the projection of $\mathbf{k}_1$ on the plane of the QW.
Since both excitons have the same energy, the emission has a maximum at
symmetric angles, i.e. when $\theta_1 = \theta_2$ (see Fig.~\ref{fig:symm_ent}a).

The dependence of entanglement on the direction of the outgoing photons and
on the parameters of the excitation pulse is relatively simple in this case.
Generally, the maximum entanglement is reached near $\theta=0$ for the
linear polarization where the non-normalized eigenvalues are $\rho_{1,2} =
[1\pm \cos(\beta)]^2$. For linearly polarized excitation field,
$\beta=\pi/2$, the pairs are maximally entangled $E_N=1$ while with
decreasing the degree of ellipticity entanglement decreases. As the special
feature the flat maximum near $\theta = 0$ should be emphasized. There
$\partial E/\partial \theta \propto \sin^3(\theta)$ for arbitrary
polarization of the excitation field. Moreover, for linearly polarized light
when $\chi=\pi/2$ entanglement reaches the maximum, $E_N=1$, and is
independent of $\theta$ (see Fig.~\ref{fig:symm_ent}b). It should be noted,
however, that as follows from Eq.~\eqref{eq:eigen_values_symmetric} the
eigenvalues of the reduced density matrix are $\rho_{1,2} \propto
\cos^2(\theta)$ in this case, so the signal vanishes in the direction
$\theta=\pi/2$.

Using Eq.~\eqref{eq:two_particle_polarization} in the polarization basis the
states are written as $\ket{\Psi} =
\sum_{\boldsymbol{\epsilon},\boldsymbol{\epsilon}'}
\Psi_{\boldsymbol{\epsilon},\boldsymbol{\epsilon}'}\ket{{\boldsymbol{\epsilon},\boldsymbol{\epsilon}'}}$,
where $\Psi_{\boldsymbol{\epsilon},\boldsymbol{\epsilon}'}$ are found
convoluting the polarization vectors of the outgoing photons
$\boldsymbol{\epsilon}$ and $\boldsymbol{\epsilon}'$ with
$\tensor{M}_{\widehat{k}_1,\widehat{k}_2}$. Along the direction $\theta
\approx 0$, where $E_N=1$ can be reached, we obtain
\begin{equation}\label{eq:maximally_ent_hh}
  \ket{\Psi} \propto -e^{i\chi}\left[1-\cos(\beta)\right]\ket{+}\ket{+}
  -e^{-i\chi}\left[1+\cos(\beta)\right]\ket{-}\ket{-},
\end{equation}
where $\ket{+}$ ($\ket{-}$) is the state of a photon that is right (left)
circularly polarized. As $\theta$ is increased, the entanglement is reduced
down except for the case $\beta = \pi/2$, $\chi = \pi/2$, where the state is
of the same structure as for $\theta = 0$, i.e. $\ket{\Psi} \propto
\ket{+}\ket{+}-\ket{-}\ket{-}$, as long as $\theta < \pi/2$. Along
$\theta=\pi/2$ the two-photon state is $\ket{\Psi} \propto
[\cos(\chi)-i\cos(\beta)\sin(\chi)]\ket{s}\ket{s}$, i.e. the two-photon
state is completely disentangled. Here the two-photon state is expressed in
terms of the $s$ and $p$ polarization eigenstates. Varying the ellipticity
of the incoming pulse away from linear polarization $\beta=\pi/2$, the
entanglement is monotonously reduced down to $E_N=0$ for $\beta = 0$. There
the two-photon state reads $\ket{\Psi} = \ket{\theta_1}\ket{\theta_1}$
with
$\ket{\theta_1}=\left(i\ket{s}+\cos(\theta)\ket{p}\right)/\sqrt{1+\cos^2(\theta)}$,
which is also completely disentangled.
We would like to emphasize that all these states are triplet, that is
transform according to the 3D representation of the rotation group.

Our calculations including the microcavity show that entanglement near $\theta = 0 $
is not affected by the cavity. However, in general its angular dependence
becomes highly non-trivial, approximately following that of
transmittivities, as is illustrated in Fig.~\ref{fig:symm_ent}c.
These properties are very useful for tailoring the angle dependence of the emission of the entangled photons.
The details of the cavity model will be discussed
elsewhere.

{\it For the photons in resonance with the light-hole excitons} ($\omega_1 =
\omega_2 = E_l$) the directional dependence of entanglement is more complex.
The reason is the interaction of obliquely propagating $p$ polarized photons
with the light hole excitons with the zero projection of the total spin. The
characteristic equation in this case differs from
Eq.~\eqref{eq:eigen_values_symmetric} by the term $X\left\{X\left[\rho - 1 +
\sin^2(\beta)\sin^2{\chi}\right] + 2\cos(\chi)\cos^2(\theta)\left[\rho -
\sin^2(\beta)\right]\right\}$ in the right-hand side with
$X=8\sin(\beta)\sin^2(\theta)$.
This term becomes important for oblique directions thus yielding a richer
structure of the angular dependence of entanglement, as shown in
Fig.~\ref{fig:symm_ent}d.

%

{\it For the photons in resonance with heavy- and light-hole excitons},
corresponding to $\theta_1 \ne\theta_2$, there is only one resonance,
when the two-exciton state is made of light-hole and heavy-hole excitons.
Respectively, there are only two resonant directions with $\theta_1
\ne\theta_2$ with
\begin{equation}\label{eq:nonsym_directions}
  \sin(\theta_\pm) = \sin(\theta_1) \left(\frac{E_l}{\Delta_{hl}}\right)^{\pm
  1}.
\end{equation}
The reduced single-photon density matrix is determined by
Eq.~\eqref{eq:reduced_density_matrix} with
$\overleftrightarrow{M}_{\widehat{\mathbf{k}}_1,\widehat{\mathbf{k}}_2} \propto A_+A_-
 \left(
 \overleftrightarrow{1}- \widehat{\mathbf{e}}_z \otimes \widehat{\mathbf{e}}_z\right)$.
Its non-normalized eigenvalues are $1$
and $\cos^2(\theta)$. As a result, entanglement monotonously decreases
from $1$ to $0$  as $\theta$ changes from $0$ to $\pi/2$ while the direction
of the detection of the second photon is determined by
Eq.~\eqref{eq:nonsym_directions}. For sufficiently large detection angles
$\theta$ such that $\sin(\theta)> \Delta_{hl}/E_l$ only one resonant
direction corresponding to $\theta_-$ remains. Interestingly, in
the asymmetric case entanglement is independent of the the polarization of
the excitation field, however,
$\overleftrightarrow{M}_{\widehat{\mathbf{k}}_1,\widehat{\mathbf{k}}_2}$ vanishes if the
excitation pulse is circularly polarized.


Studying entanglement would not be complete without considering yield, which
is defined as the ratio of the energy flux carried by the entangled pairs of
the photons to the flux of the excitation field. For practical purposes it
is more convenient to use an alternative definition $Y= N_{out}/N_{in}$,
where $N_{out}$ and $N_{in}\sim  (\Phi/\hbar \Omega)^2$ are the number of
outgoing and incoming \textit{pairs}, respectively, with $\Phi$ being the
total flux of the external field and $\Omega$ being its frequency in the
stationary frame.
Calculating the trace of the single-particle
density matrix over the polarization quantum numbers we find (in SI units)
\begin{equation}\label{eq:yield}
  Y \sim \frac{4\pi^9 E_x}{3\gamma^3 S}(\rho_1 + \rho_2)
  \left(\frac{T |Q|^2}{\hbar \Omega c^3 \epsilon_0^2}\right)^2,
\end{equation}
where $E_x$ is the exciton energy, $\rho_{1,2}$ are the non-normalized
eigenvalues of the density matrix defined by the equations studied above,
$S$ is the area of the excitation spot, $T$ is the duration of the
excitation pulse, $Q$ is the common interband dipole moment, and
$\epsilon_0$ is vacuum permittivity. The inverse dependence on the pump area
is clear since the dense excitation more pronounced is the effect of the
exclusion principle. The dependence on the pulse duration is the consequence
of the semiconductor response determined by the polarizations of the linear
response rather than by the energy input.
Substituting the values typical for GaAs, $\gamma = 1.5$ meV, $\hbar
\Omega \approx E_x = 1.5$ eV, $Q = 1.3\cdot 10^{-24}$
$\mathrm{kg}\cdot\mathrm{m}/\mathrm{s}$ (see e.g.
Ref.~\onlinecite{Yu_Cardona}) and using $T=100$ fs, $S = \pi (20)^2$
$\mu\mathrm{m}^2$ we find $Y \sim 0.02$. Such high value of yield is the
result of the resonant transitions between the many-particle states.


In conclusion, we have studied the basic mechanism of emission entangled
photons by a semiconductor quantum well excited by a short pulse. We have
developed a ``kinematic" theory accounting the effect of the exclusion
principle.
The dependence of entanglement on the detection angle and on the
polarization of the external field is shown to be highly nontrivial. We have
estimated yield of the considered process and have found it to be rather
high owing to the resonant transitions between different states.


\begin{thebibliography}{20}
\expandafter\ifx\csname natexlab\endcsname\relax\def\natexlab#1{#1}\fi
\expandafter\ifx\csname bibnamefont\endcsname\relax
  \def\bibnamefont#1{#1}\fi
\expandafter\ifx\csname bibfnamefont\endcsname\relax
  \def\bibfnamefont#1{#1}\fi
\expandafter\ifx\csname citenamefont\endcsname\relax
  \def\citenamefont#1{#1}\fi
\expandafter\ifx\csname url\endcsname\relax
  \def\url#1{\texttt{#1}}\fi
\expandafter\ifx\csname urlprefix\endcsname\relax\def\urlprefix{URL }\fi
\providecommand{\bibinfo}[2]{#2} \providecommand{\eprint}[2][]{\url{#2}}

\bibitem[{\citenamefont{Shih and Alley}(1988)}]{Shih}
    \bibinfo{author}{\bibfnamefont{Y.~H.} \bibnamefont{Shih}}
    \bibnamefont{and}
  \bibinfo{author}{\bibfnamefont{C.~O.} \bibnamefont{Alley}},
  \bibinfo{journal}{Phys.~Rev.~Lett.} \textbf{\bibinfo{volume}{61}},
  \bibinfo{pages}{2921} (\bibinfo{year}{1988}).

\bibitem[{\citenamefont{Ou and Mandel}(1988)}]{Ou}
    \bibinfo{author}{\bibfnamefont{Z.~Y.} \bibnamefont{Ou}}
    \bibnamefont{and}
  \bibinfo{author}{\bibfnamefont{L.}~\bibnamefont{Mandel}},
  \bibinfo{journal}{Phys.~Rev.~Lett.} \textbf{\bibinfo{volume}{61}},
  \bibinfo{pages}{50} (\bibinfo{year}{1988}).

\bibitem[{\citenamefont{Altepeter et~al.}(2005)\citenamefont{Altepeter,
  Jeffrey, and Kwiat}}]{Altepeter}
\bibinfo{author}{\bibfnamefont{J.~B.} \bibnamefont{Altepeter}},
  \bibinfo{author}{\bibfnamefont{E.~R.} \bibnamefont{Jeffrey}},
  \bibnamefont{and} \bibinfo{author}{\bibfnamefont{P.~G.} \bibnamefont{Kwiat}},
  \bibinfo{journal}{Optics Express} \textbf{\bibinfo{volume}{13}},
  \bibinfo{pages}{8951} (\bibinfo{year}{2005}).

\bibitem[{\citenamefont{Kwiat et~al.}(1995)\citenamefont{Kwiat, Mattle,
  Weinfurter, Zeilinger, Sergienko, and Shih}}]{Kwiat}
\bibinfo{author}{\bibfnamefont{P.~G.} \bibnamefont{Kwiat}},
  \bibinfo{author}{\bibfnamefont{K.}~\bibnamefont{Mattle}},
  \bibinfo{author}{\bibfnamefont{H.}~\bibnamefont{Weinfurter}},
  \bibinfo{author}{\bibfnamefont{A.}~\bibnamefont{Zeilinger}},
  \bibinfo{author}{\bibfnamefont{A.~V.} \bibnamefont{Sergienko}},
  \bibnamefont{and} \bibinfo{author}{\bibfnamefont{Y.}~\bibnamefont{Shih}},
  \bibinfo{journal}{Phys.~Rev.~Lett.} \textbf{\bibinfo{volume}{75}},
  \bibinfo{pages}{4337} (\bibinfo{year}{1995}).

\bibitem[{\citenamefont{Bouwmeester et~al.}(2000)\citenamefont{Bouwmeester,
  Ekert, and Zeilinger}}]{Bouwmeester}
\bibinfo{author}{\bibfnamefont{D.}~\bibnamefont{Bouwmeester}},
  \bibinfo{author}{\bibfnamefont{A.~K.} \bibnamefont{Ekert}}, \bibnamefont{and}
  \bibinfo{author}{\bibfnamefont{A.}~\bibnamefont{Zeilinger}},
  \emph{\bibinfo{title}{The Physics of Quantum Information}}
  (\bibinfo{publisher}{Springer}, \bibinfo{address}{Berlin},
  \bibinfo{year}{2000}).

\bibitem[{\citenamefont{Mandel and Wolf}(1995)}]{Mandel}
    \bibinfo{author}{\bibfnamefont{L.}~\bibnamefont{Mandel}}
    \bibnamefont{and}
  \bibinfo{author}{\bibfnamefont{E.}~\bibnamefont{Wolf}},
  \emph{\bibinfo{title}{Optical Coherence and Quantum Optics}}
  (\bibinfo{publisher}{Cambridge University Press}, \bibinfo{address}{New
  York}, \bibinfo{year}{1995}).

\bibitem[{\citenamefont{Tanzilli et~al.}(2001)\citenamefont{Tanzilli,
  Riedmatten, Tittel, Zbinden, Baldi, Micheli, Ostrowski, and
  Gisin}}]{Tanzilli}
\bibinfo{author}{\bibfnamefont{S.}~\bibnamefont{Tanzilli}},
  \bibinfo{author}{\bibfnamefont{H.~D.} \bibnamefont{Riedmatten}},
  \bibinfo{author}{\bibfnamefont{W.}~\bibnamefont{Tittel}},
  \bibinfo{author}{\bibfnamefont{H.}~\bibnamefont{Zbinden}},
  \bibinfo{author}{\bibfnamefont{P.}~\bibnamefont{Baldi}},
  \bibinfo{author}{\bibfnamefont{M.~D.} \bibnamefont{Micheli}},
  \bibinfo{author}{\bibfnamefont{D.~B.} \bibnamefont{Ostrowski}},
  \bibnamefont{and} \bibinfo{author}{\bibfnamefont{N.}~\bibnamefont{Gisin}},
  \bibinfo{journal}{Electronics Letters} \textbf{\bibinfo{volume}{37}},
  \bibinfo{pages}{26} (\bibinfo{year}{2001}).

\bibitem[{\citenamefont{Hafenbrak et~al.}(2007)\citenamefont{Hafenbrak,
    Ulrich,
  Michler, Wang, Rastelli, and Schmidt}}]{Hafenbrak}
\bibinfo{author}{\bibfnamefont{R.}~\bibnamefont{Hafenbrak}},
  \bibinfo{author}{\bibfnamefont{S.~M.} \bibnamefont{Ulrich}},
  \bibinfo{author}{\bibfnamefont{P.}~\bibnamefont{Michler}},
  \bibinfo{author}{\bibfnamefont{L.}~\bibnamefont{Wang}},
  \bibinfo{author}{\bibfnamefont{A.}~\bibnamefont{Rastelli}}, \bibnamefont{and}
  \bibinfo{author}{\bibfnamefont{O.~G.} \bibnamefont{Schmidt}},
  \bibinfo{journal}{New Journal of Physics} \textbf{\bibinfo{volume}{9}},
  \bibinfo{pages}{315} (\bibinfo{year}{2007}).

\bibitem[{\citenamefont{Stevenson et~al.}(2006)\citenamefont{Stevenson,
    Young,
  Atkinson, Cooper, Ritchie, and Shields}}]{STEVENSON:2006:ID3501}
\bibinfo{author}{\bibfnamefont{R.~M.} \bibnamefont{Stevenson}},
  \bibinfo{author}{\bibfnamefont{R.~J.} \bibnamefont{Young}},
  \bibinfo{author}{\bibfnamefont{P.}~\bibnamefont{Atkinson}},
  \bibinfo{author}{\bibfnamefont{K.}~\bibnamefont{Cooper}},
  \bibinfo{author}{\bibfnamefont{D.~A.} \bibnamefont{Ritchie}},
  \bibnamefont{and} \bibinfo{author}{\bibfnamefont{A.~J.}
  \bibnamefont{Shields}}, \bibinfo{journal}{Nature}
  \textbf{\bibinfo{volume}{439}}, \bibinfo{pages}{179} (\bibinfo{year}{2006}).

\bibitem[{\citenamefont{Akopian et~al.}(2007)\citenamefont{Akopian, Lindner,
  Poem, Berlatzky, Avron, Gershoni, Gerardot, and Petroff}}]{Akopian2}
\bibinfo{author}{\bibfnamefont{N.}~\bibnamefont{Akopian}},
  \bibinfo{author}{\bibfnamefont{N.~H.} \bibnamefont{Lindner}},
  \bibinfo{author}{\bibfnamefont{E.}~\bibnamefont{Poem}},
  \bibinfo{author}{\bibfnamefont{Y.}~\bibnamefont{Berlatzky}},
  \bibinfo{author}{\bibfnamefont{J.}~\bibnamefont{Avron}},
  \bibinfo{author}{\bibfnamefont{D.}~\bibnamefont{Gershoni}},
  \bibinfo{author}{\bibfnamefont{B.~D.} \bibnamefont{Gerardot}},
  \bibnamefont{and} \bibinfo{author}{\bibfnamefont{P.~M.}
  \bibnamefont{Petroff}}, \bibinfo{journal}{Journal of Applied Physics}
  \textbf{\bibinfo{volume}{101}}, \bibinfo{pages}{081712}
  (\bibinfo{year}{2007}).

\bibitem[{\citenamefont{Young et~al.}(2006)\citenamefont{Young, Stevenson,
  Atkinson, Cooper, Ritchie, and Shields}}]{Young}
\bibinfo{author}{\bibfnamefont{R.~J.} \bibnamefont{Young}},
  \bibinfo{author}{\bibfnamefont{R.~M.} \bibnamefont{Stevenson}},
  \bibinfo{author}{\bibfnamefont{P.}~\bibnamefont{Atkinson}},
  \bibinfo{author}{\bibfnamefont{K.}~\bibnamefont{Cooper}},
  \bibinfo{author}{\bibfnamefont{D.~A.} \bibnamefont{Ritchie}},
  \bibnamefont{and} \bibinfo{author}{\bibfnamefont{A.~J.}
  \bibnamefont{Shields}}, \bibinfo{journal}{New Journal of Physics}
  \textbf{\bibinfo{volume}{8}}, \bibinfo{pages}{29} (\bibinfo{year}{2006}).

\bibitem[{\citenamefont{Shields}(2007)}]{Shields}
    \bibinfo{author}{\bibfnamefont{A.~J.} \bibnamefont{Shields}},
  \bibinfo{journal}{Nature Photonics} \textbf{\bibinfo{volume}{1}},
  \bibinfo{pages}{215} (\bibinfo{year}{2007}).

\bibitem[{\citenamefont{Edamatsu et~al.}(2004)\citenamefont{Edamatsu,
    Oohata,
  Shimizu, and Itoh}}]{EDAMATSU:2004:ID3421}
\bibinfo{author}{\bibfnamefont{K.}~\bibnamefont{Edamatsu}},
  \bibinfo{author}{\bibfnamefont{G.}~\bibnamefont{Oohata}},
  \bibinfo{author}{\bibfnamefont{R.}~\bibnamefont{Shimizu}}, \bibnamefont{and}
  \bibinfo{author}{\bibfnamefont{T.}~\bibnamefont{Itoh}},
  \bibinfo{journal}{Nature} \textbf{\bibinfo{volume}{431}},
  \bibinfo{pages}{167} (\bibinfo{year}{2004}).

\bibitem[{\citenamefont{Oohata et~al.}(2007)\citenamefont{Oohata, Shimizu,
    and
  Edamatsu}}]{Oohata}
\bibinfo{author}{\bibfnamefont{G.}~\bibnamefont{Oohata}},
  \bibinfo{author}{\bibfnamefont{R.}~\bibnamefont{Shimizu}}, \bibnamefont{and}
  \bibinfo{author}{\bibfnamefont{K.}~\bibnamefont{Edamatsu}},
  \bibinfo{journal}{Phys.~Rev.~Lett.} \textbf{\bibinfo{volume}{98}},
  \bibinfo{pages}{140503} (\bibinfo{year}{2007}).

\bibitem[{\citenamefont{Savasta et~al.}(2003)\citenamefont{Savasta,
    Di~Stefano,
  and Girlanda}}]{SAVASTA:2003_SST}
\bibinfo{author}{\bibfnamefont{S.}~\bibnamefont{Savasta}},
  \bibinfo{author}{\bibfnamefont{O.}~\bibnamefont{Di~Stefano}},
  \bibnamefont{and} \bibinfo{author}{\bibfnamefont{R.}~\bibnamefont{Girlanda}},
  \bibinfo{journal}{Semicond. Sci. Technol.} \textbf{\bibinfo{volume}{18}},
  \bibinfo{pages}{S294} (\bibinfo{year}{2003}).

\bibitem[{\citenamefont{Kira et~al.}(1999)\citenamefont{Kira, Jahnke, Hoyer,
  and Koch}}]{KIRA:1999}
\bibinfo{author}{\bibfnamefont{M.}~\bibnamefont{Kira}},
  \bibinfo{author}{\bibfnamefont{F.}~\bibnamefont{Jahnke}},
  \bibinfo{author}{\bibfnamefont{W.}~\bibnamefont{Hoyer}}, \bibnamefont{and}
  \bibinfo{author}{\bibfnamefont{S.}~\bibnamefont{Koch}},
  \bibinfo{journal}{Prog. Quant. Electr.} \textbf{\bibinfo{volume}{23}},
  \bibinfo{pages}{189} (\bibinfo{year}{1999}).

\bibitem[{\citenamefont{Shurkliff}(1962)}]{Shurkliff}
    \bibinfo{author}{\bibfnamefont{W.}~\bibnamefont{Shurkliff}},
  \emph{\bibinfo{title}{Polarized Light}} (\bibinfo{publisher}{Harvard
  University Press}, \bibinfo{address}{Cambridge}, \bibinfo{year}{1962}).

\bibitem[{\citenamefont{Yu and Cardona}(2004)}]{Yu_Cardona}
    \bibinfo{author}{\bibfnamefont{P.}~\bibnamefont{Yu}} \bibnamefont{and}
  \bibinfo{author}{\bibfnamefont{M.}~\bibnamefont{Cardona}},
  \emph{\bibinfo{title}{Fundamentals of Semiconductors: Physics and Materials
  Properties}} (\bibinfo{publisher}{Springer}, \bibinfo{address}{Berlin},
  \bibinfo{year}{2004}).

\bibitem[{\citenamefont{Ostreich et~al.}(1998)\citenamefont{Ostreich,
  Schonhammer, and Sham}}]{OSTREICH:1998}
\bibinfo{author}{\bibfnamefont{T.}~\bibnamefont{Ostreich}},
  \bibinfo{author}{\bibfnamefont{K.}~\bibnamefont{Schonhammer}},
  \bibnamefont{and} \bibinfo{author}{\bibfnamefont{L.~J.} \bibnamefont{Sham}},
  \bibinfo{journal}{Phys. Rev. B} \textbf{\bibinfo{volume}{58}},
  \bibinfo{pages}{12920} (\bibinfo{year}{1998}).

\bibitem[{\citenamefont{Scully and Zubairy}(1997)}]{Scully_QO}
    \bibinfo{author}{\bibfnamefont{M.}~\bibnamefont{Scully}}
    \bibnamefont{and}
  \bibinfo{author}{\bibfnamefont{M.}~\bibnamefont{Zubairy}},
  \emph{\bibinfo{title}{Quantum optics}} (\bibinfo{publisher}{Cambridge
  University Press}, \bibinfo{address}{Cambridge}, \bibinfo{year}{1997}).

\end{thebibliography}

\section{Supplemental information}

In the lowest non-vanishing order of the perturbation theory Since the
tensor $\overleftrightarrow{M}_{\widehat{k}_1,\widehat{k}_2}(t)$ does not
depend on the polarization states of the outgoing photons we omit the hats
over $k$'s and write it solely in terms of the wave vectors of the outgoing
photons
\begin{equation}\label{eq:M_tensor}
\begin{split}
 \overleftrightarrow{M}_{\mathbf{k}_1,\mathbf{k}_2}(t) = \frac{2\pi}{4\sqrt{\omega_{{k}_1}\omega_{{k}_2}}}
 \widetilde{\phi}_{\mu_1}(0) \widetilde{\phi}_{\mu_2}(0)
  \int_0^t & dt_1 e^{-i (\omega_{{k}_1} - \omega_{{k}_2})(t-t_1)}\int_0^{t_1} dt_2\,
  e^{-iE_{\mu_1}(t_1-t_2)}\\
  & \left[e^{-i\omega_{{k}_2}(t_1-t_2)}
  \delta(\mathbf{K}_{\mu_1}-\mathbf{k}_{1,\|})\delta(\mathbf{K}_{\mu_2}-\mathbf{k}_{2,\|})
  \,
  \mathbf{d}_{\sigma_{\mu_1},s_{\mu_1}} \otimes \mathbf{d}_{\sigma_{\mu_2},s_{\mu_2}}
  \right. \\
  &  \left.
   +
  e^{-i\omega_{{k}_1}(t_1-t_2)}
  \delta(\mathbf{K}_{\mu_2}-\mathbf{k}_{1,\|})\delta(\mathbf{K}_{\mu_1}-\mathbf{k}_{2,\|})
  \,
  \mathbf{d}_{\sigma_{\mu_2},s_{\mu_2}} \otimes \mathbf{d}_{\sigma_{\mu_1},s_{\mu_1}}
  \right]\mathcal{G}_{\mu_1,\mu_2}(t_2),
\end{split}
\end{equation}
where $\otimes$ denotes the tensor product so that in a particular
coordinate system $(\mathbf{d}\otimes \mathbf{d}')_{ij} = d_i d'_j$,
$\omega_k = k$ is the photon energy and $E_\mu$ are the energies of the
hole-electron pair states. The effect of the semiconductor nonlinear
response is described by the function $\mathcal{G}_{\mu_1,\mu_2}(t)$, which
we  present as a sum of the instantaneous term and the memory term
\begin{equation}\label{eq:G_separation}
 \mathcal{G}_{\mu_1,\mu_2}(t) =
 \aver{B_{\mu_1}B_{\mu_2}B^\dagger_{\nu_3}B^\dagger_{\nu_4}}P^{(1)}_{\nu_3}(t)P^{(1)}_{\nu_4}(t)
 +i \int_0^t dt'\, \aver{B_{\mu_1}B_{\mu_2} e^{-iH_{SC}(t-t')}D^\dagger_{\nu_3, \nu_4}}P^{(1)}_{\nu_3}(t')P^{(1)}_{\nu_4}(t'),
\end{equation}
where\cite{OSTREICH:1998} $D^\dagger_{\nu_3, \nu_4} =
[B^\dagger_{\nu_3},[B^\dagger_{\nu_4},H]]$ and $P^{(1)}_{\nu}(t)$ are the
exciton polarizations of the linear response created by the action of the
external (classical) field
\begin{equation}\label{eq:linear_response}
 P^{(1)}_{\nu}(t) = -i \int_0^t dt'\, e^{-iE_\nu(t-t')}\mathcal{A}^{(cl)}_\nu(t').
\end{equation}

Neglecting the memory term (or the contribution of the Coulomb interaction)
the tensor $\overleftrightarrow{M}_{\mathbf{k}_1,\mathbf{k}_2}(t)$ is
directly expressed in terms of $A_+$ and $A_-$, the amplitudes of the left
circular and right circular components of the excitation pulse,
respectively,

\begin{equation}\label{eq:tensor_M_reduced}
\begin{split}
 \overleftrightarrow{M}_{\mathbf{k}_1,\mathbf{k}_2}(t) = \frac{(2\pi)^5}{4\sqrt{\omega_1\omega_2}} \,\delta(\mathbf{k}_{1, \|} + \mathbf{k}_{2, \|})
& \left(A_-^2\widehat{\mathbf{e}}_+\otimes\widehat{\mathbf{e}}_+
 +
 A_+^2\widehat{\mathbf{e}}_-\otimes\widehat{\mathbf{e}}_-\right)
 \left[W_{hh}(t) + \frac 1 9 W_{ll}(t)\right]
 \\
 &+ \frac 2 3 \left(\widehat{\mathbf{e}}_+\otimes\widehat{\mathbf{e}}_-
 + \widehat{\mathbf{e}}_-\otimes\widehat{\mathbf{e}}_+\right)A_+A_-
 \left[W_{hl}(t) + W_{lh}(t)\right] \\
 & - \frac 8 9 \widehat{\mathbf{e}}_z\otimes\widehat{\mathbf{e}}_z A_+A_-W_{ll}(t).
\end{split}
\end{equation}
Here the indices $h$ and $l$ denote $|\sigma| = 3/2$ and $|\sigma| = 1/2$,
respectively. The time dependence is described by $W_{\sigma_1,\sigma_2}(t)
= \mathcal{I}_{\sigma_1,\sigma_2} \left[w_{\sigma_1,\sigma_2}(\omega_1,
\omega_2; t) + w_{\sigma_1,\sigma_2}(\omega_2, \omega_1; t)\right]$ with
\begin{equation}\label{eq:w_definition}
  w_{\sigma_1,\sigma_2}(\omega_1, \omega_2; t) = \int_0^t dt_1 e^{-i(\omega_1 + \omega_2)(t-t_1)}
  \int_0^{t_1}dt_2 e^{-i(\omega_1 + E_x)(t_1-t_2)}e^{-i E_{xx}t_2},
\end{equation}
where $E_{x} = E_{\sigma_1}$ and $E_{xx} = E_{\sigma_1} + E_{\sigma_2}$
denote single-exciton and two-exciton energies, respectively. The intensity
of the exciton-light interaction is quantified by $\mathcal{I}_{\sigma_1,
\sigma_2} =
  |Q|^4 \left|\widetilde{\phi}_{\sigma_1}(0) \widetilde{\phi}_{\sigma_2}(0)\right|^2
  \int d\mathbf{q}\,
  \left|\widetilde\varphi_{\sigma_1}(\mathbf{q})\right|^2
  \left|\widetilde\varphi_{\sigma_2}(\mathbf{q})\right|^2$ with $\widetilde\varphi_{\sigma}(\mathbf{q})$ being the Fourier transform of
the exciton wave function. Here and in Eq.~\eqref{eq:tensor_M_reduced} we
have taken into account the structure of the valence band in the
semiconductors with the point symmetry $T_d$ and have introduced the common
interband dipole moment $Q = -i e/m \bra{X}p_x\ket{\Gamma_1}$ (see e.g.
Ref.~\onlinecite{Yu_Cardona}).

The important result immediately following from
Eq.~\eqref{eq:tensor_M_reduced} is that if the external excitation field is
circularly polarized then the emitted photons are disentangled. Indeed, in
this case the tensor $\overleftrightarrow{M}_{\mathbf{k}_1,\mathbf{k}_2}(t)$
is represented as a tensor product and, hence, so is the single-particle
density matrix, i.e. it corresponds to a pure state.

Representation~\eqref{eq:tensor_M_reduced} explicitly shows the tensor
$\overleftrightarrow{M}_{\mathbf{k}_1,\mathbf{k}_2}(t)$ as a superposition
of the amplitudes of the radiative decay of different two-exciton states
through two channels into the two-photon states. Due to the presence of
different characteristic frequencies the time dependence of the total
amplitude has a complex form, especially at the transitional regime with the
duration proportional to inversed heavy-hole light-hole splitting. However,
there are several resonances whose amplitudes increases with time and which
define the long time response and respectively the time dependence of
entanglement. In order to extract these resonances we consider the long time
limit in the spirit of the Wigner-Weisskopf approximation \cite{Scully_QO}
extending the limits of integrations over time to infinity. This yields
\begin{equation}\label{eq:m_typical}
  w(t) \propto e^{-iE_{xx} t} \delta\left[E_{xx}(1+\alpha - b) - E_{x}(1+\alpha)\right]
 \delta\left[\omega_1 + \omega_2 -E_{xx}\right].
\end{equation}
Here we have taken into account the momentum selection rule and have
introduced $\alpha = \sin(\theta_1)/\sin(\theta_2)$ and $b$ stands for
either $1$ or $\alpha$ for $w(\omega_1,\omega_2)$ and
$w(\omega_2,\omega_1)$, respectively, that is depending on particular
exciton-photon channel. As follows from Eq.~\eqref{eq:m_typical} only such
terms in Eq.~\eqref{eq:tensor_M_reduced} contribute into the long limit
which satisfy the special resonant condition. The total energy of the
emitted pair must be equal to the energy of the two-exciton state.
Additionally there is the special ``kinematic" requirement imposed on the
energies of the involved single- and two-exciton states.

\end{document}